\input harvmac
\input epsf

\def \strace{ \mathop{ \rm strace}\nolimits}
\def \I{ \mathop{ \rm I}\nolimits}
\def \J{ \mathop{ \rm J}\nolimits}
\def \sech{ \mathop{ \rm sech}\nolimits}
\def \Re{ \mathop{ \rm Re}\nolimits}
\def \Im{ \mathop{ \rm Im}\nolimits}
\Title{ DFTUZ /98/10}
{\vbox{\centerline{EXACT SOLUTION OF A ELECTRON SYSTEM }
\vskip2pt\centerline{COMBINING TWO DIFFERENT $t\hbox{-}J$ MODELS}}}
\centerline{J. Abad \footnote{$^{(1)}$}{julio@posta.unizar.es}
and M. R\'{\i}os\footnote{$^{(2)}$}{mrios@encomix.es} }
\centerline{Departamento de F\'{\i}sica Te\'{o}rica, Facultad de Ciencias,}
\centerline{Universidad de Zaragoza, 50009 Zaragoza, Spain}
\bigskip
\bigskip
\vskip .3in
\centerline{ \tenbf Abstract}
A new strongly correlated electron model is presented. This is formed 
by two types of sites: one where double occupancy is forbidden, as 
in the $t\hbox{-}J$ model, and the other where double occupancy is 
allowed but vacancy is not allowed, as an inverse $t\hbox{-}J$ model. The 
Hamiltonian shows nearest and next-to-nearest neighbour interactions 
and it is solved by means of a modified algebraic nested Bethe Ansatz. The 
number of sites where vacancy is not allowed, may be treated as a new 
parameter if the model is looked at as a $t\hbox{-}J$ model with impurities.  The ground and excited states are described in the thermodynamic limit.

\noindent PACS number: 75.10.Jm

\vskip .2in
\bigskip
\bigskip
\Date{}

\vfill
\eject

\newsec{Introduction}

Strongly correlated systems are very interesting in view of their relation 
with high $T_{c}$ superconductivity. It is very important to study the 
one-dimensional systems because they may share properties with 
two-dimensional ones \ref\ri{ P.W. Anderson, {\it Phys. Rev. Lett.} {\bf 65} (1990) 
2306.}.  The $t\hbox{-}J$ model was proposed by F.C. Zhang 
and T.M. Rice \ref\rii {  F.C. Zhang and T.M. Rice, {\it Phys. Rev.} {\bf B65} 
(1988) 3759.} and it describes electrons on a lattice excluding 
the double occupancy of any site, in opposition to the Hubbard model 
\ref\riii{J. Hubbard, {\it Proc. Roy. Soc.} {\bf 276} 
	(1963) 238. {\it Proc. Roy. Soc.} {\bf 277} (1964) 237.
	{\it Proc. Roy. Soc.} {\bf 284} (1964) 401.}, where the double occupancy is not forbidden. At the point $J=2t$ 
the model (called supersymmetric) is associated to a graded Lie algebra and 
it is exactly integrable. Their integrability was studied in 
\ref\riv{  P. Schlottmann, {\it Phys. Rev.} {\bf B36} 
	(1987) 5177.},  
although equivalent systems had been solved by other authors as in
\ref\rv{C.K. Lai, {\it J. Math. Phys.} {\bf 15} (1974) 1675.} and
\ref\rvb{ B. Sutherland, {\it Phys. Rev.} {\bf B12} (1975) 3795.}.
 The ground state and the excitation spectrum were investigated in 
\ref\rvi{  P.A. Bares and G. Blatter, {\it Phys. Rev. Lett.} {\bf 
	64} (1990) 2567.},\ref\rvib{ P.A. Bares, G. Blatter and M. Ogata, {\it Phys. 
	Rev.} {\bf B44} (1991) 130.} and 
low-lying excitations close to half-filling were treated in \ref\rvii{S. Sarkar, {\it J. Phys.} {\bf A23} 
	(1990) L409. {\it J. Phys.} {\bf A24} (1991) 1137. }. The 
integration of the model by means of the nested Algebraic Bethe Ansatz (NABA)
\ref\rviii{ P.P. Kulish and N.Y. Reshetikhin, {\it J. 
	Phys.} {\bf A16} (1983) L591.
},\ref\rviiib{ L. Takhtajan, {\it J. Soviet Math.} {\bf 23} 
	(1983) 2470.} in the framework of the Graded Quantum Inverse Scattering Method 
(GQISM) \ref\rix{P. Kulish and E.K. Sklyanin, {\it J. Soviet Math.} {\bf 19} (1982) 1596.
\hfill\break
P. Kulish, {\it J. Soviet Math.} {\bf 35} (1986) 2648.} was established in \ref\rx{ F.H.L. E{\ss}ler and V.E. Korepin, {\it Phys. Rev.} {\bf 
	B46} (1992) 9147.}. The completeness of the 
Bethe states was considered in \ref\rxi{ A. Foerster and M. Karowski, {\it Phys. Rev.} {\bf 
	B46} (1992) 9234. {\it Nucl. Phys.} {\bf B396} (1993) 611.} and the properties of the model 
in an external magnetic field were studied in \ref\rxii{ P.A. Bares, J.M.P. Carmelo, J. Ferrer and P. Horsch, 
	{\it Phys. Rev.} {\bf B46} (1992) 14624.}.

In this paper we propose a model with two different kinds of sites. The 
sites of the first type may be unoccupied or occupied by an electron with spin 
up or down, but the double occupancy is not allowed. We will call these 
sites as $t\hbox{-}J$. The sites of the second type may be occupied by an 
electron with spin up or down, or by two electrons with antiparallel spins 
in a singlet state, but vacancies are not possible. We will call 
these sites as \lq \lq$J\hbox{-}t$" because it is as an inverse $t\hbox{-}J$ model. We can 
look  these sites as frustrated Hubbard sites. Other $t\hbox{-}J$ models with impurities have been proposed by different authors \ref\rxiin{ P. A. Bares, " Exact results for one dimensional $t\hbox{-}J$ model with impurities", cond-mat/9412011.
\hfill\break
G. Berd\"{u}rftig, F. H. L. Essler and H. Frahm, {\it Nucl. Phys.} {\bf B498} (1997) 697. }

In order to make the inhomogeneus system we have used the $R$-matrix of the $t\hbox{-}J$ model and 
the Graded Yang-Baxter Equation (GYBE). We have take other solution that we
associate to the "$J\hbox{-}t$" sites, that solution can give us another integrable  homogeneus system. The system that we propose is formed by alternating $N_h$ states in the $t\hbox{-}J$ sites with $N_{p}$ states in the  $J\hbox{-}t$ sites.

As we will see, the method is trivially dependent of the number of  sites of each class, then we are going to diagonalized the transfer matrix for a general lattice with 
$N_h$  and $N_{p}$ states in the respectives sites, although we have used an 
alternating chain with $N_h=N_{p}=N/2$ to compute the Hamiltonian.  This Hamiltonian is the sum 
of nearest neighbor interaction terms (two site operators) and 
next-to-nearest neighbor interaction terms (three-site operators).  The diagonalization is made using the method that we proposed in ref. \ref \rxvsn{J. Abad and M. R\'{\i}os, {\it Phys. Rev.} {\bf B53} (1996) 14000. {\it J. Phys.} {\bf A 30} (1997) 5887.}. This method is more general than the usual NABA, and it has been used by other authors in a different system \ref \nnn { M. P. Pfannmuller and H. Frahm, {\it J. Phys. A: Math. Gen.} {\bf 30} (1997) L 543.}. Recently J. Links and A. Foerster have proposed a model \ref\nnni {J. Links and A. Foerster,  \lq \lq Integrability of a $t\hbox{-}J$ model with impurities" , cond-mat/9806129. } with the same kind of alternating site states that in our model and they solve it  with the same method. 

 As we said before and will see along this article,  all resuls that we have, are independent of the  number and position of the two kinds of sites, then our system can be considered as a $t\hbox{-}J$ model with impurities $J\hbox{-}t$.

\newsec{The model}

The $t\hbox{-}J$ model may be deduced, in the framework of the graded Lie 
algebras, by means of the $L$-operator. We are going to take the operators given by E{\ss}ler and  Korepin in ref. \rx\ .
\eqn\eai{
L^{(t-J)}(\lambda) = \lambda I + i P  ,
}
where
\eqna\eaii
$$\eqalignno{
	& I^{a,b}_{c,d} = \delta_{a,b} \delta_{c,d} &\eaii a \cr
	& P^{a,b}_{c,d} = \delta_{a,d} \delta_{b,c} (-1)^{\epsilon_{b} 
	 \epsilon_{d}}   &\eaii b \cr}
$$
and $\epsilon_{j}$ are the Grassmann parities of the basis vectors. 
We will use in this paper the FFB basis, that is, $\epsilon_{1}=\epsilon_{2}=1$ 
and $\epsilon_{3}=0$. The upper indices are in the space that  we call auxiliary, and the lower indices in the site space.

The operator $L^{(t-J)}$ verifies the GYBE
\eqn\eaiii{
	R(\lambda-\mu)\biggl[L^{(t-J)}(\lambda) \otimes L^{(t-J)}(\mu)\biggr]= 
	\biggl[L^{(t-J)}(\mu) \otimes L^{(t-J)}(\lambda)\biggr] 
	R(\lambda-\mu),
	}
where the $R$-matrix is given by
\eqn\eaiv{
	R(\lambda) = \lambda P + i I,
	}
and the tensor product is the graded tensor product, which is defined 
as
\eqn\eav{
	(F \otimes G)^{a,b}_{c,d} = F_{a,b}G_{c,d}(-1)^{\epsilon_{c} 
	(\epsilon_{a}+\epsilon_{b})}.
}
Equation \eaiii\ in components is,
\eqnn\eaiiin
$$\eqalignno{
	&R(\lambda-\mu)^{a,c_1}_{b,d_1}L^{(t-J)}(\lambda)^{c_1,c}_{g,e_1} L^{(t-J)}(\mu)_{e_1,h}^{d_1,d} (-1)^{\epsilon_{d_1}(\epsilon_{c_1}+\epsilon_{c})}= \cr 
&\qquad\qquad\qquad\qquad\qquad	=L^{(t-J)}(\mu)^{a,c_2}_{g,e_2}L^{(t-J)}(\lambda)^{b,d_2}_{e_2,h}
	R(\lambda-\mu)^{c_2,c}_{d_2,d} (-1)^{\epsilon_{b}(\epsilon_{a}+\epsilon_{c_2})}
	&\eav\cr
}$$

In order to make a mixed lattice, we need other $L$-operator 
associated to the $J\hbox{-}t$ sites. That operator must  fulfil the same GYBE eq.  \eaiii\  . Inspired in the same methods given in ref. \rx\ , we have found that
\eqn\eavi{ 
	L^{(J-t)}(\lambda) = (\lambda-{ i\over 2}) I -i Q
}
verifies our requirements, being
\eqn\eavii{
    Q^{a,b} _{c,d}= \delta_{a,c} \delta_{b,d} (-1)^{\epsilon_{a} \epsilon_{c}}.
}
We have used the $t\hbox{-}J$ basis
\eqn\eaviii{
| 0  \rangle =\left (\matrix{
0\cr
0\cr
1\cr
}\right ), \qquad
| \downarrow \rangle =
\left (\matrix{
0\cr
1\cr
0\cr
}\right ), \qquad
| \uparrow \rangle = \left (\matrix{
1\cr
0\cr
0\cr
}\right ),
}
and the $J\hbox{-}t$ basis
\eqn\eaix{
|  \downarrow \uparrow\rangle =\left (\matrix{
0\cr
0\cr
1\cr
}\right ), \qquad
| \downarrow \rangle =
\left (\matrix{
0\cr
1\cr
0\cr
}\right ), \qquad
| \uparrow \rangle = \left (\matrix{
1\cr
0\cr
0\cr
}\right ),
}

It is easy to show by direct calculation
\eqn\eax{
	L^{(J-t)}(\lambda)^{a,b}_{\alpha,\beta}
	L^{(J-t)}(-\lambda)^{\beta,\gamma}_{b,c} = \rho(\lambda)
	\delta_{a,c}\delta_{\alpha,\gamma}
}
with
\eqn\eaxi{
	\rho(\lambda) = -{1 \over 4}-\lambda^{2}.
}
%
%
%
%
%

We build the monodromy matrix alternating the $t\hbox{-}J$ and the $J\hbox{-}t$ 
L-operators. 
\eqnn\eaxii
$$\eqalignno{
 T_{a,b}(\lambda,w) &=
	 L^{(t-J)}(\lambda)^{a,c_{1}}_{a_{1},b_{1}}
	 L^{(J-t)}(\lambda+w)^{c_{1},c_{2}}_{\alpha_{1},\beta_{1}}
	 L^{(t-J)}(\lambda)^{c_{2},c_{3}}_{a_{2},b_{2}} \cdots
	 \cr
       & \qquad \qquad \cdots 
	  L^{(t-J)}(\lambda)^{c_{N-2},c_{N-1}}_{a_{N/2},b_{N/2}}
	 L^{(J-t)}(\lambda+w)^{c_{N-1},b}_{\alpha_{N/2},\beta_{N/2}} \times
 \cr
	 & \times (-1)^{\left[
	 \sum_{i=1}^{N/2}(a_{i}+b_{i})\sum_{j=1}^{N/2}(\alpha_{j}+a_{j+1})+
	 \sum_{i=1}^{N/2-1}(\alpha_{i}+\beta_{i})\sum_{j=1}^{N/2-1}
	 (a_{j+1}+\alpha_{j+1})
	 \right]},
&\eaxii \cr}
$$
with $N$ even and $a_{j}=\alpha_{j}=0$ if $j>N/2$.
We have used the standard matrix product in the auxiliary 
space and the graded tensorial product in the site space.

In the auxiliary space we can write the monodromy matrix as
\eqn\eaxiii{
T(\lambda,w) =\pmatrix{
A_{1,1}(\lambda,w) & A_{1,2}(\lambda,w) & B_{1}(\lambda,w)  \cr
A_{2,1}(\lambda,w) & A_{2,2}(\lambda,w) & B_{2}(\lambda,w)  \cr
C_{1}(\lambda,w) & C_{2}(\lambda,w) & D(\lambda,w)\cr
} 
} 
The transfer matrix is given by 
\eqn\eaxiv{
F(\lambda,w)=\strace (T(\lambda,w)) =  D(\lambda,w) -
	A_{1,1}(\lambda,w) - A_{2,2}(\lambda,w),
}
and the verification of (GYBE) by the monodromy matrix assures the 
commutation of the transfer matrices for different values of the 
argument, that is
\eqn\eaxv{
	\left[
	F(\lambda,w),F(\mu,w)
	\right] = 0.
}

The corresponding associate Hamiltonian is obtained by taking the first 
logarithmic derivative of the transfer matrix at equal zero spectral parameter
\eqn\eaxvi{
	H(w) = -i J {d \over{d\lambda}} \ln 
	(F(\lambda,w)) \big|_{\lambda=0},
}
where $J$ is a constant.

We write this Hamiltonian as the sum both operators: the nearest 
neighbor interaction term and the next-to-nearest neighbor 
interaction term
\eqn\eaxvii{
	H(w) = {{-iJ} \over \rho(w)}\sum_{j=1 \atop j \, odd}^{N-1}
	h^{[1]}_{j,j+1} + {{-iJ} \over c_{1}\rho(w)}
	\sum_{j=1 \atop j \, odd}^{N-1}	h^{[2]}_{j,j+1,j+2}\, ,
}
with $R(0) = c_{1}I$ and
\eqna \eaxviii
$$\eqalignno{
& \left(h^{[1]}_{j,j+1}\right)_{a,b;\beta,\gamma} = 
	 \dot{L}^{(J-t)}(w)^{a,c}_{\beta,\delta}
	  L^{(J-t)}(-w)^{\delta,\gamma}_{c,b}
&\eaxviii a \cr
	 & \left(h^{[2]}_{j,j+1,j+2}\right)_{a,b;\beta,\gamma;c,d} = 
	 L^{(J-t)}(w)^{a,e}_{\beta,\delta}
	 \dot{L}^{(t-J)}(0)^{e,d}_{c,f}
	  L^{(J-t)}(-w)^{\delta,\gamma}_{f,b}
	  (-1)^{\epsilon_{c}(\epsilon_{\beta}+\epsilon_{\delta})}. 
&\eaxviii b\cr	 
}
$$
The dot over the operator shows the derivative.
%
%
%
%
%
%
%

Taking $w=0$ we find that,
\eqnn \eaxix
$$\eqalignno{
 H=& J \sum_{j=1 \atop j \, odd}^{N-1}\biggl\{-2\biggl\{
	 \sum_{\sigma=\uparrow,\downarrow}^{}
	 (c_{j,\sigma}d_{j+1,\sigma}-h.c.) - 2n_{j}\bar{n}_{j+1} 
\cr
 & +\sum_{\sigma=\uparrow,\downarrow}^{}
	 (d_{j+1,\sigma}c_{j+2,\sigma}-h.c.) - 2\bar{n}_{j+1}n_{j+2} \biggr\} - 6n_{j}-4\bar{n}_{j+1} \cr
& +3 \biggl\{
	 \sum_{\sigma=\uparrow,\downarrow}^{}(c^{\dag}_{j,\sigma}c_{j+2,\sigma}+
	 h.c.) +(S^{-}_{j}S^{+}_{j+2}+S^{+}_{j}S^{-}_{j+2}+2S^{z}_{j}S^{z}_{j+2})
	 \biggr\} \cr
& -{11 \over 2}n_{j}n_{j+2}- 2\biggl\{
	 \sum_{\sigma=\uparrow,\downarrow}^{}\epsilon(\sigma)\biggl[
	 (S^{z}_{j}d^{\dag}_{j+1,\sigma}c^{\dag}_{j+2,\sigma}-
	 c^{\dag}_{j,\sigma}S^{z}_{j+1}c_{j+2,\sigma} \cr
& -c^{\dag}_{j,\sigma}d^{\dag}_{j+1,\sigma}S^{z}_{j+2}) + h.c.
	 \biggr] \biggr\} +
	 3\biggl\{\biggl[
	 \sum_{\sigma=\uparrow,\downarrow}^{}
	 (c_{j,\sigma}d_{j+1,\sigma}-3 h.c.)  \cr
& -2({1 \over 4}n_{j}\bar{n}_{j+1}+S^{-}_{j}S^{-}_{j+1}+ 
	 S^{+}_{j}S^{+}_{j+1} + 2S^{z}_{j}S^{z}_{j+1})
	 \biggr]\cdot n_{j+2}\biggr\}   \cr
& + n_{j}\cdot \bigg[\sum_{\sigma=\uparrow,\downarrow}^{}
	 (d_{j+1,\sigma}c_{j+2,\sigma} - 3 h.c.) + 2({1 \over 4}
	 \bar{n}_{j+1}n_{j+2} +S^{-}_{j+1}S^{-}_{j+2}  \cr
& +S^{+}_{j+1}S^{+}_{j+2} + 2S^{z}_{j+1}S^{z}_{j+2})
	 \bigg]   
	 -2\biggl\{\sum_{\sigma=\uparrow,\downarrow}^{}\bigg[
	 (S^{\tau(\sigma)}_{j}d_{j+1,\sigma}c_{j+2,\bar{\sigma}}\cr
&  -c_{j,\sigma}S^{\tau(\sigma)}_{j+1}c_{j+2,\bar{\sigma}}-
	 c^{\dag}_{j,\sigma}d^{\dag}_{j+1,\bar{\sigma}}S^{\tau(\sigma)}_{j+2})
	 + h.c.
	 \biggr]\biggr\}  \cr
& -\biggl\{3\sum_{\sigma=\uparrow,\downarrow}^{}
	 (c^{\dag}_{j,\sigma}\bar{n}_{j+1}c_{j+2,\sigma} + h.c.) +
	 2(S^{-}_{j}\bar{n}_{j+1}S^{+}_{j+2}  \cr
 &  +S^{+}_{j}\bar{n}_{j+1}S^{-}_{j+2}+
	 2 S^{z}_{j}\bar{n}_{j+1}S^{z}_{j+2})
	 \biggr\}\biggr\} + 3N &\eaxix \cr
}$$
where $c^{\dag}$ and $n$ are the creation operator and the number of 
electrons operator in the $t\hbox{-}J$ sites, and $d^{\dag}$ and $\bar{n}$ 
are the same operators in the $J\hbox{-}t$ sites. Also,
$S$ is the spin operator.  In terms of the elements of the $L^{t\hbox{-}J}$ operators, they are

\eqnn\eaxxa
$$\eqalignno{
&{[}c _{\uparrow} ^{\dagger}{]} _{a,b} ={1 \over i}L^{(t-J)} ( \lambda  )^{3,1}_{a,b}  \cr
&{[}c _{\downarrow} ^{\dagger}{]} _{a,b}={1 \over i}L^{(t-J)}( \lambda ) ^{3,2}_{a,b}     \cr
&{[}c _{\uparrow} {]} _{a,b} ={1 \over i}L ^{(t-J)}( \lambda )^{1,3}_{a,b}            \cr
&{[}c _{\downarrow}{]} _{a,b}={1 \over i}L^{(t-J)}( \lambda )^{2,3}_{a,b}     \cr
&{[}S ^{+}{]} _{a,b}= - {1 \over i}L^{(t-J)}( \lambda )^{2,1}_{a,b}               \cr
&{[}S ^{-}{]} _{a,b}=-{1 \over i}L^{(t-J)}( \lambda )^{1,2}_{a,b}    \cr
&{[}S ^{z}{]} _{a,b}={i \over 2}
 \left\{ L^{(t-J)}(\lambda )^{1,1}_{a,b}-L^{(t-J)}(\lambda )^{2,2}_{a,b}\right\}  
 \cr
&{[}n{]} _{a,b}=-i{ \lambda +i\over \lambda -i}
 \left\{ L^{(t-J)}(\lambda )^{1,1}_{a,b}+L^{(t-J)}(\lambda )^{2,2}_{a,b}\right\} +
{ 2(\lambda ^{2}-i\lambda +1)\over \lambda (i\lambda +1)}
L ^{(t-J)}( \lambda )
 ^{3,3}_{a,b}  ,             &\eaxxa \cr
}$$

and in terms of the $L^{J\hbox{-}t}$ 

\eqnn\eaxxib
$$\eqalignno{
&{[}d _{\uparrow} ^{\dagger}{]} _{a,b}={1 \over i}L^{(J-t)}( \lambda )^{1,3}_{a,b} \cr
&{[}d _{\downarrow} ^{\dagger}{]} _{a,b}={1 \over i}L^{(J-t)}( \lambda )^{2,3}_{a,b}  \cr
&{[}d _{\uparrow}{]} _{a,b}=-{1 \over i}L^{(J-t)}( \lambda )^{3,1}_{a,b}   \cr
&{[}d _{\downarrow}{]} _{a,b}=-{1 \over i}L^{(J-t)}( \lambda )^{3,2}_{a,b}     \cr
&{[}S ^{+}{]} _{a,b}={1 \over i}L ^{(J-t)}( \lambda )^{1,2}_{a,b}            \cr
&{[}S ^{-}{]} _{a,b}={1 \over i}L^{(J-t)}( \lambda )^{2,1}_{a,b}     \cr
&{[}S ^{z}{]} _{a,b}=-{i \over 2}\left\{ L^{(J-t)}(\lambda)^{1,1}_{a,b}-L^{(J-t)}(\lambda)^{2,2}_{a,b}\right\}  \cr
&{[} \overline{n}{]} _{a,b}=-i
 \left\{ L^{(J-t)}(\lambda )^{1,1}_{a,b}+L^{(J-t)}(\lambda )^{2,2}_{a,b}\right\} -
{ {2\lambda -3i}\over {2\lambda -3i}}
L ^{(J-t)}( \lambda )^{3,3}_{a,b}   &\eaxxib \cr
}$$

 We have also used the following notation,
\eqna \eaxx
$$\eqalignno{
& \bar{\sigma}=\cases{
\downarrow & if  $\sigma = \uparrow $ \cr
\uparrow & if  $\sigma =\downarrow$ \cr
}
&\eaxx a  \cr
& \tau(\uparrow) = \downarrow\,, \quad
	 \tau(\downarrow) = \uparrow\,, \qquad
	 \epsilon(\uparrow) = 1\,, \quad
	 \epsilon(\downarrow) = -1\,. \qquad
	&\eaxx b \cr 
}$$
Obviously, the Hamiltonian \eaxix\  is not hermitian, however as we will see in section IV, it has real eigenvalues. In refs. \ref\rxiii{A. Foerster and M. Karowski, {\it Nucl. 
	Phys.} {\bf B408} (1993) 512. } and \ref\rxiv{ A. Gonz\'{a}lez-Ruiz, {\it Nucl. 
	Phys.} {\bf B424} (1994) 468. }.is show that  the supersymmetric $t\hbox{-}J$ no hermitian hamiltonians with quantum group invariance enjoy this property. In a most general case, in ref. \ref\rxiiia {C. M. Bender and S Boettcher, {\it Phys. Rev. Lett. } {\bf 80 } (1998) 5243.} complex hamiltonians with $PT$ invariance is proof that they have real spectra. 

\newsec{Algebraic Bethe Ansatz}

The monodromy operator $T$ verifies the GYBE, independently of the combination of  ${t\hbox{-}J}$ and ${J\hbox{-}t}$ sites that we take in our system. Then, we are going to take $N_h$ sites of the first type and  $N_p$ of the second. The space of states of the total system will be,
\eqn\ebni{
E=  \bigotimes^{N_{h}}_{i} E_i ^{t\hbox{-}J}\bigotimes^{N_{p}}_{j} E_j^{J\hbox{-}t}
}
where $E_i ^{t\hbox{-}J}$ is the space of states of the site $i$  and $E_j^{J\hbox{-}t} $ the space of states  of the site $j$.

In order to diagonalize the Hamiltonian by means of the Nested Algebraic Bethe 
Ansatz (NABA), we should need to find a state of the system that verifies \rix ,
\eqn\ebi{
	A_{i,j} \Vert v \rangle \propto \delta_{i,j}\Vert v \rangle,
}
but this is not the case. To overcome the problem, we must follow a modified Nested Algebraic Bethe Ansatz 
MNABA, which is described in reference \rxvsn. For this purpose, in a first step, we build the vacuum subspace 
\eqn\ebii{
	\Omega = \bigotimes^{N_{h}}_{i} | 0 {\rangle}_{i}
	\bigotimes^{N_{p}}_{j} \{e\}_{j}\,,
}
where $\{e\}_j$ denotes the subspace generated by the vectors $|\downarrow 
\rangle$ and $| \uparrow \rangle$ in a $J\hbox{-}t$ site $j$. Any $ \Vert w {\rangle}$
vector in the vacuum subspace verifies
\eqna \ebiii
$$\eqalignno{
& A_{i,j}(\lambda) \Vert w {\rangle} \in \Omega \qquad , \qquad i,j=1,2
	&\ebiii a\cr
& C_{i}(\lambda) \Vert w {\rangle} \neq 0 \qquad , \qquad i=1,2
	&\ebiii b  \cr
& B_{i}(\lambda) \Vert w {\rangle} = 0 \qquad , \qquad i=1,2
	&\ebiii c  \cr
& D(\lambda) \Vert w {\rangle} = \left[a'(\lambda)\right]^{N_{0}}
	 \left[b_{1}(\lambda)\right]^{N_{p}} \Vert w {\rangle} 
         &\ebiii d  \cr
}$$
where $\Vert w {\rangle} \in \Omega$ and 
\eqn\ebiv{	a'(\lambda) = \lambda + i\,, \qquad b_{1}(\lambda) = \lambda 
	-{1 \over 2}.
}

From the GYBE we get the relation of commutation
\eqna \ebv
$$\eqalignno{
& A_{a,b}(\mu)C_{c}(\lambda) = (-1)^{\epsilon_{a}\epsilon_{p}} 
	 g(\mu-\lambda) r(\mu-\lambda)^{d,c}_{p,b} C_{p}(\lambda) A_{a,d}(\mu)
&\    \cr
& \qquad \qquad \qquad + h(\mu-\lambda) C_{b}(\mu) A_{a,c}(\lambda),
	&\ebv a \cr
& D(\mu)C_{c}(\lambda) = g(\lambda-\mu)C_{c}(\lambda)D(\mu) - 
	 h(\lambda-\mu)C_{c}(\mu)D(\lambda), \qquad 
	&\ebv b  \cr
& C_{a}(\lambda)C_{b}(\mu) = r(\lambda-\mu)^{d,b}_{c,a} 
	 C_{c}(\mu)C_{d}(\lambda) ,
	&\ebv c\cr
}$$
where 
\eqna \ebvi
$$\eqalignno{
& g(\mu) = {\mu+i \over \mu},
	&\ebvi a  \cr
& h(\mu) ={ i \over \mu},
	&\ebvi b\cr
& r(\mu)^{a,b}_{c,d} = {h(\mu) \over g(\mu)} \delta_{a,b} 
	 \delta_{c,d} - {1 \over g(\mu)} \delta_{a,d} \delta_{b,c}  .
&\ebvi c \cr
}$$

In order to solve the equation
%
%
%
%
\eqn\evi
{
	F(\lambda) \Psi = \Lambda(\lambda) \Psi ,
}
we build the state, 
\eqn\ebviii{
	\Psi(\vec{\lambda}) \equiv \Psi(\lambda_{1},\ldots,\lambda_{r}) =
	C_{a_{1}}(\lambda_{1}) \ldots C_{a_{r}}(\lambda_{r})
	X^{a_{1}\ldots a_{r}} \Vert 1 {\rangle},
}
with $ \Vert 1 {\rangle} \in \Omega$.
When we apply the $D(\mu)$ operator to $\Psi$, using\ebv{a-c}, we push this operator to the right of the $C$ operators, and we get a wanted term characterized by the $C$ operators conserve their arguments, and several unwanted terms characterized by the arguments of the $C$ operators are interchanged. The wanted term is,
\eqn\ebix{
	\left[a'(\mu)\right]^{N_{0}} \left[b_{1}(\mu)\right]^{N_{p}}
	\prod_{j=1}^{r} g(\lambda_{j}-\mu) \Psi,
}
and the $k$-th unwanted term is,
\eqnn \ebx
$$\eqalignno{
-h(\lambda_{k}-\mu) 
	&\left[a'(\lambda_{k})\right]^{N_{0}} \left[b_{1}(\lambda_{k})\right]^{N_{p}}
	\prod_{j=1}^{r} g(\lambda_{j}-\lambda_{k})
	C_{b_{k}}(\mu) C_{b_{k+1}}(\lambda_{k+1}) \ldots 
	  \cr
&\qquad \qquad \ldots C_{b_{k-1}}(\lambda_{k-1}) 
	 M^{(k-1)}(\lambda_{k-1})^{b_{1},\ldots,b_{r}}_{a_{1},\ldots,a_{r}}
	 X^{a_{1},\ldots,a_{r}} \Vert 1 {\rangle}, &\ebx \cr
}$$
with $M$ given in the appendix A.
%
%

The application of $A_{a,a}$ to the state $\Psi$ is a little larger but straightforward. We get again wanted and unwanted terms, and after some calculations we find the wanted term,
\eqn\ebxi{
	\prod_{j=1}^{r}g(\mu-\lambda_{j}) C_{p_{1}}(\lambda_{1}) \ldots 
	C_{p_{r}}(\lambda_{r}) A_{a,b}(\mu) 
	\left[Z(\mu,\vec{\lambda})^{p_{1},\ldots,p_{r}}_{a_{1}, 
	\ldots,a_{r}}\right]_{b,a} X^{a_{1},\ldots,a_{r}} \Vert 1 {\rangle}.
}
where the $Z$ operator is
\eqnn \ebxii
$$\eqalignno{
 &\left[Z(\mu,\vec{\lambda})^{p_{1},\ldots,p_{r}}_{a_{1}, 
	\ldots,a_{r}}\right]_{i,j} = 
	L^{(1)}(\mu-\lambda_{r})^{i,d_{r-1}}_{p_{r},a_{r}}
	L^{(1)}(\mu-\lambda_{r-1})^{d_{r-1},d_{r-2}}_{p_{r-1},a_{r-1}} \ldots	\qquad   \cr
& \qquad \qquad \qquad \ldots
	 L^{(1)}(\mu-\lambda_{2})^{d_{2},d_{1}}_{p_{2},a_{2}}
	L^{(1)}(\mu-\lambda_{1})^{d_{1},j}_{p_{1},a_{1}} \,,
	&\ebxii\cr
}$$
with
\eqn\ebxiii{
	L^{(1)}(\lambda)^{f,d}_{e,b} = r(\lambda)^{a,b}_{c,d}
	P^{(1)}{}^{e,a}_{f,c},
}
and $P^{(1)}{}^{e,a}_{f,c} = \delta_{e,c}\delta_{a,f} 
(-1)^{\epsilon_{a}\epsilon_{c}}$.

Due to $j=1$ or $j=2$, we have $\epsilon_{j}=1$, then \ebxiii\
becomes

\eqn\ebxiv{
	L^{(1)}(\lambda)^{f,d}_{e,b} = -r(\lambda)^{f,b}_{e,d}\,.
}

Now we will get ready for prepare the second step of the MNABA.
We define the monodromy matrix at second level as
\eqn\ebxv{
T^{(2)}(\mu,\vec{\lambda}) =A(\mu) \cdot Z(\mu,\vec{\lambda})\,.
}
In the auxiliary space, that now his two-dimensional,  this operator can be written as
\eqn\ebxvi{
T^{(2)}(\mu,\vec{\lambda}) =\pmatrix{
A^{(2)}(\mu,\vec{\lambda})  & C^{(2)}(\mu,\vec{\lambda})  \cr
B^{(2)}(\mu,\vec{\lambda})  & D^{(2)}(\mu,\vec{\lambda}) \cr
}
}
The transfer matrix at second level is
\eqn\ebxvii{
	F^{(2)}(\mu,\vec{\lambda}) = \strace \left[
	T^{(2)}(\mu,\vec{\lambda}) \right] =
	- A^{(2)}(\mu,\vec{\lambda}) -D^{(2)}(\mu,\vec{\lambda}).
}
Now, the wanted term \ebxi\ can be written as,
\eqn\ebxviii{
	-\prod_{j=1}^{r} g(\mu-\lambda_{j}) C_{p_{1}}(\lambda_{1}) \ldots 
	C_{p_{r}}(\lambda_{r}) 
	F^{(2)}(\mu,\vec{\lambda})^{p_{1},\ldots,p_{r}}_{a_{1},\ldots,a_{r}}
	X^{a_{1},\ldots,a_{r}} \Vert 1 {\rangle},
}
and the $k$-th unwanted term as
\eqnn\ebxix
$$\eqalignno{
h(\mu-\lambda_{k})& \prod_{j=1 \atop j \neq k}^{r} 
	 g(\lambda_{k}-\lambda_{j}) C_{p_{k}}(\mu)
	 C_{p_{k+1}}(\lambda_{k+1}) \ldots C_{p_{r}}(\lambda_{r})
	 C_{p_{1}}(\lambda_{1}) \ldots
\cr
& \ldots C_{p_{k-1}}(\lambda_{k-1})
	 M^{(k-1)}(\lambda_{k-1})^{p_{1},\ldots,p_{r}}_{b_{1},\ldots,b_{r}}
	 F^{(2)}(\lambda_{k},\vec{\lambda})^{b_{1},\ldots,b_{r}}_{a_{1}, 
	 \ldots,a_{r}} X^{a_{1},\ldots,a_{r}} \Vert 1 {\rangle}.
&\ebxix\cr
}$$
We have found a new problem; we have to solve the equation,
\eqn\ebxx{
	F^{(2)}(\mu,\vec{\lambda}) X \Vert 1 {\rangle} = 
	\Lambda^{(2)}(\mu,\vec{\lambda}) X \Vert 1 {\rangle}\,.
}
For this purpose we use the fact that $L^{(1)}(\lambda)$ verifies the 
GYBE with the $r$-matrix, and then we can solve \ebxx\ following 
the same path applied before.

The new vacuum is
\eqn\ebxxi{
	\Vert 1' {\rangle} =  \bigotimes^{N}_{i} | 0 \downarrow {\rangle}_{i}
	\bigotimes^{r}_{j} 
\left (\matrix{
1\cr
0\cr
}\right ),
}
where the state  $| 0 \downarrow {\rangle}$ is composed by a state $| 0 {\rangle}$ in the $t\hbox{-}J$ sites and $| \downarrow 
{\rangle}$ in the $J\hbox{-}t$ sites. The first tensorial product is the 
space where $A(\mu)$ works, whereas $Z(\mu,\vec{\lambda})$ works in 
the second one. This vacuum verifies
\eqna\ebxxii
$$\eqalignno{
& B^{(2)}(\mu,\vec{\lambda}) \Vert 1' {\rangle} = 0 \quad ,
	&\ebxxii a  \cr
& C^{(2)}(\mu,\vec{\lambda}) \Vert 1' {\rangle} \neq 0\quad ,
	&\ebxxii b\cr
& A^{(2)}(\mu,\vec{\lambda}) \Vert 1' {\rangle} = \left[
	 b(\mu)\right]^{N_{0}} \left[b_{1}(\mu)\right]^{N_{p}} 
	 \prod_{j=1}^{r} {g(\lambda_{j}-\mu) \over g(\mu-\lambda_{j})}
	 \Vert 1' {\rangle}\quad ,
	&\ebxxii c\cr
& D^{(2)}(\mu,\vec{\lambda}) \Vert 1' {\rangle} = \left[
	 b(\mu)\right]^{N_{0}} \left[a_{1}(\mu)\right]^{N_{p}} 
	 \prod_{j=1}^{r} {1 \over g(\mu-\lambda_{j})} \Vert 1' {\rangle}\quad ,
	&\ebxxii d\cr
}$$
with
\eqn\ebxxiii{
	a_{1}(\lambda) = \lambda +{i \over 2} \,, \qquad b(\lambda) = \lambda \,.
}
The new relations of commutation are obtained from the GYBE for 
$T^{(2)}$
\eqna\ebxxiv
$$\eqalignno{
& D^{(2)}(\mu) C^{(2)}(\lambda) = g(\lambda-\mu) C^{(2)}(\lambda)
	  D^{(2)}(\mu) +  h(\mu-\lambda)  C^{(2)}(\mu) D^{(2)}(\lambda) ,\quad
	  \qquad	&\ebxxiv a  \cr
& A^{(2)}(\mu) C^{(2)}(\lambda) = g(\mu-\lambda) C^{(2)}(\lambda)
	  A^{(2)}(\mu) + h(\lambda-\mu)C^{(2)}(\mu) A^{(2)}(\lambda) \,. \qquad
	  \quad &\ebxxiv b  \cr
}$$
In the second level, we build the state
\eqn\ebxxv{
	X\Vert 1 {\rangle} \equiv \Psi^{(2)}(\mu_{1},\ldots,\mu_{s}) =
	C^{(2)}(\mu_{1}) \ldots C^{(2)}(\mu_{s}) \Vert 1' {\rangle} \,.
}
Applying $A^{(2)}$ to this state we get a wanted term
\eqn\ebxxvi{
	\left[b(\mu)\right]^{N_{0}} \left[b_{1}(\mu)\right]^{N_{p}}
	\prod_{i=1}^{s} g(\mu-\mu_{i}) \prod_{j=1}^{r}
	{g(\lambda_{j}-\mu) \over g(\mu-\lambda_{j})} \Psi^{(2)},
}
and  unwanted terms of which, the $k$-th is
\eqnn\ebxxvii
$$\eqalignno{
h(\mu_{k}-\mu)
	 \left[b(\mu_{k})\right]^{N_{0}}& \left[b_{1}(\mu_{k})\right]^{N_{p}}
	 \prod_{i=1 \atop i \neq k}^{s} g(\mu_{k}-\mu_{i}) \prod_{j=1}^{r}
	{g(\lambda_{j}-\mu_{k}) \over g(\mu_{k}-\lambda_{j})}
	  \qquad \qquad
	\cr
& \qquad \qquad \times 
	 C^{(2)}(\mu) C^{(2)}(\mu_{k+1}) \ldots C^{(2)}(\mu_{k-1})
	 \Vert 1' {\rangle} \,.
	&\ebxxvii\cr
}$$
Also for $D^{(2)}$ we get a wanted and unwanted, the wanted is,
\eqn\ebxxviii{
	\left[b(\mu)\right]^{N_{0}} \left[a_{1}(\mu)\right]^{N_{p}}
	\prod_{i=1}^{s} g(\mu_{i}-\mu) \prod_{j=1}^{r}
	{1 \over g(\mu-\lambda_{j})} \Psi^{(2)},
}
and the $k$-th unwanted term is,
%
%
%
\eqnn\ebxxix
$$\eqalignno{
h(\mu-\mu_{k})&
	 \left[b(\mu_{k})\right]^{N_{0}} \left[a_{1}(\mu_{k})\right]^{N_{p}}
	 \prod_{i=1 \atop i \neq k}^{s} g(\mu_{i}-\mu_{k}) \prod_{j=1}^{r}
	{1 \over g(\mu_{k}-\lambda_{j})}
	  \qquad \qquad
	\cr
& \qquad \qquad \times 
	 C^{(2)}(\mu) C^{(2)}(\mu_{k+1}) \ldots C^{(2)}(\mu_{k-1})
	 \Vert 1' {\rangle} \,.
	&\ebxxix\cr
}$$

The cancellation of the unwanted terms, \ebx\ with \ebxix\ and 
\ebxxvii\ with \ebxxix , gives us the ansatz equations
\eqna\ebxxx
$$\eqalignno{
& \left[{b(\lambda_{k}) \over a'(\lambda_{k})}\right]^{N_{0}}
	 = \prod_{i=1}^{s} {1 \over g(\lambda_{k}-\mu_{i})} \,,\qquad \qquad 
	 k=1,\ldots,r
	&\ebxxx a  \cr
& \left[{a_{1}(\mu_{n}) \over b_{1}(\mu_{n})}\right]^{N_{p}}
	 = \prod_{j=1}^{r} g(\lambda_{j}-\mu_{n})
	 \prod_{i=1 \atop i\neq n}^{s}
	 {g(\mu_{n}-\mu_{i}) \over g(\mu_{i}-\mu_{n})}\,,\quad
	 n=1,\ldots,s.
&\ebxxx b\cr
}$$
On the other hand, collecting the wanted terms we get the eigenvalue of the transfer matrix,
\eqnn\ebxxxi
$$\eqalignno{
\Lambda(\mu) &=
 	 \left[a'(\mu)\right]^{N_{0}} \left[b_{1}(\mu)\right]^{N_{p}}
 	 \prod_{j=1}^{r} g(\lambda_{j}-\mu) - \left[b(\mu)\right]^{N_{0}} 
 	 \times
 	\cr 
& \left\{
 	 \left[b_{1}(\mu)\right]^{N_{p}} 
 	 \prod_{i=1}^{s} g(\mu-\mu_{i}) \prod_{j=1}^{r} g(\lambda_{j}-\mu) +
 	 \left[a_{1}(\mu)\right]^{N_{p}} 
 	 \prod_{i=1}^{s} g(\mu_{i}-\mu)
 	 \right\}.
 	&\ebxxxi\cr
 }$$


\newsec{Thermodynamics of the model}

The eigenstates of the transfer matrix  can be characterized by observables operators  that commute with it, and then with the Hamiltonian. We have found that the following observables
\eqna \eci
$$\eqalignno{
&O_1=\sum_{i=1 \atop i \, odd}^{N-1} (n_i+\bar{n}_{i+1}), &\eci a\cr
&O_2=\sum_{i=1 \atop i \, odd}^{N-1} (n_i^{holes}-\bar{n}_{i+1}^{pairs} ),&\eci b\cr
&O_3=\sum_{i=1 \atop i \, odd}^{N-1} (n_{i,\uparrow}+\bar{n}_{i+1,\downarrow} ),&\eci c\cr
&O_4=\sum_{i=1 \atop i \, odd}^{N-1} (n_{i,\downarrow}+\bar{n}_{i+1,\uparrow} ),&\eci d\cr
}$$
commute  with the transfer matrix
\eqn\ecii{
[O_i, F(\lambda)]=0, \qquad i=1,\dots, 4.
}
We must note that $O_1$  is the total number of the electrons, $O_2$ is the difference between the number of holes and number of electron pairs, $O_3$ is the number of electrons with the spin up in the $\{ |0>, |\downarrow>,|\uparrow> \}$ base plus the number of electrons with the spin down in the $\{ |\downarrow, \uparrow>, |\downarrow>,|\uparrow> \}$ base and $O_4$ is the complimentary of $O_3$. Only two of these quantities are independent, the others are linearly dependents and thus, we are going to use $O_1$ and $O_3$.

The commutation relations of the $C_i$ creation operators with these observables allows us to relation their eigenvalues with the numbers $r$ and $s$ of the ansatz equations. They are:
\eqna \eciii
$$\eqalignno{
&[O_1, C_i]= C_i, &\eciii a\cr
&[O_2, C_i]= -C_i, &\eciii b\cr
&[O_3, C_i]= \delta_{1,i} C_i, &\eciii c\cr
&[O_4, C_i]= \delta_{2,i} C_i, &\eciii d\cr
}$$
and a long but straighforward calculation, gives us the action of these operators on the eigenstate $\psi (\lambda_1,\cdots,\lambda_r)$,
\eqna\eciv
$$\eqalignno{
&O_1 \psi (\lambda_1,\cdots,\lambda_r)=(r+N_p) \psi (\lambda_1,\cdots,\lambda_r)
&\eciv a\cr
&O_2 \psi (\lambda_1,\cdots,\lambda_r)=(-r+N_h) \psi (\lambda_1,\cdots,\lambda_r)
&\eciv b\cr
&O_3 \psi (\lambda_1,\cdots,\lambda_r)=(r-s+N_p) \psi (\lambda_1,\cdots,\lambda_r)
&\eciv c\cr
&O_4 \psi (\lambda_1,\cdots,\lambda_r)=s \psi (\lambda_1,\cdots,\lambda_r)
&\eciv d\cr
}$$

The next step is to solve the equations of the ansatz 
in the thermodynamic limit,  i.e., when $N$, $N_h$ and $N_p$ are infinity, but $N_h/N $ and $N_p /N$ remain finite. For this, it is convenient to reparametrize the roots as follows
\eqna\ecriv
$$\eqalignno{
&\lambda_j = v_j^{(1)}-{i \over 2}, \qquad j=1,\cdots,r &\ecriv a\cr
&\mu_k = v_k^{(2)}, \qquad k=1,\cdots,s &\ecriv b\cr
}$$
and then the equations \ebxxx {a,b}
are written
\eqna\ecrv
$$\eqalignno{
&\biggl[{ v_k^{(1)}-{i \over 2}\over v_k^{(1)}+{i \over 2}} \biggr]^{N_h}=
\prod_{j=1}^{s}{{ v_k^{(1)}-v_j^{(2)}-{i \over 2}\over v_k^{(1)}-v_j^{(2)}+{i \over 2}} }
&\ecrv a\cr
&\biggl[{ v_k^{(2)}+{i \over 2}\over v_k^{(2)}-{i \over 2}} \biggr]^{N_p}=
-\prod_{j=1}^{r}{{ v_k^{(2)}-v_j^{(1)}-{i \over 2}\over v_k^{(2)}-v_j^{(1)}+{i \over 2}} }
\prod_{l=1}^{s}{{ v_k^{(2)}-v_l^{(2)}+i\over v_k^{(2)}-v_l^{(2)}-i} }
&\ecrv b\cr
}$$
The energy is obtained with 
\eqn\ecrvi{
E= -i J {d \over d\lambda}{\ln{\Lambda(\lambda)}}\biggl|_{\lambda=0} =-i J 
\biggl[ N_h {\dot{a}'(0) \over a'(0)}+N_p {\dot{b}_1(0) \over b_1(0)}-\sum_{j=1}^{r}{{\dot{g}(\lambda_j) \over g(\lambda_j) }} \biggr],
}
and using \ecriv\,
\eqn\ecrvii{
E=J \bigg[ -N_h +2 N_p  + \sum_{j=1}^{r}{{1 \over (v_j^{(1)})^2 +{1 \over 4}}} \biggr].
}

It is convenient to introduce the function,
\eqn\ecv{
\phi (x)\equiv 2 \arctan{(x)} =- i\ln{{1+i x \over 1-i x}},
}
and taking logarithms in the equations of the ansatz \ecrv {a,b}, 
we can write
\eqna\ecvi
$$\eqalignno{
&N_h \phi (2 v_k^{(1)}) -\sum_{j}^{s}{\phi (2 v_k^{(1)}-2v_j^{(2)})} = 2 \pi \I_k^{(1)}
&\ecvi a\cr
&N_p \phi (2 v_k^{(2)}) +\sum_{j}^{r}{\phi (2 v_k^{(2)}-2v_j^{(1)})} -\sum_{l}^{s}{\phi (v_k^{(2)}-v_l^{(1)})}= 2 \pi \I_k^{(2)}
&\ecvi b\cr
}$$
with $\I_k^{(1)}$ and $\I_k^{(2)}$ integers or half-odd integers and each set  of these numbers determines a solution for roots $v_k^{(1)}$ and $v_k^{(2)}$.

In solving a BAE set, we have  real roots and complex conjugate roots grouped in clusters that we call n-strings. The solution for the ground state in the thermodynamic limit can be obtained minimizing the free energy distribution \ref \nx{C. N. Yang and C. P, Yang, {\it Phys. Rev.} {\bf 150} (1966) 321.
\hfill\break
M. Takahashi, {\it Prog. of Th. Phys.} {\bf 46} (1971) 401.}. In the Heisenberg model with arbitrary spin is found that the solution for the ground state is formed by  two-string roots \ref\nxi{ H. M. Bubujian, {\it Nucl. Phys. B} {\bf 215} (1983) 317.}. In the $t\hbox{-}J$ model, the two level roots are mixed, they proliferate rapidly and it become difficult to determinate the roots which parametrize the ground state. Numerical analysis suggest that the structure of solutions in the ground state are two-string roots \rvib .  We assume the same hypothesis in our model. 

Then,  the solution in ground state is given by  two-string roots and for the lower excited states some roots are real.  The equations \ecvi{a,b} can  be reparametrized supposing that, in the  $r$ roots of the first level, $v_j^{(1)}$, $j=1,\dots,r$,  there are $2r_1$ roots of the form
\eqn\ecvii{
v_j^{(1)}=\epsilon_l \pm{ i \over 2} ,\qquad l=1,\cdots, r_1
}
with $\epsilon_l $ real, that are grouped in $r_1$ two-strings. The rest $r-2r_1$ roots are taken real and they are designed by,
\eqn\ecviii{
v_j^{(1)}=\eta_l ,\qquad l=1,\cdots,(r-2 r_1).
}
With the same assumptions, the $s$ roots $v_j^{(2)}$ of second level are written,
\eqnn\ecix
$$\eqalignno{
&v_j^{(2)}=\beta_l \pm{ i \over 2} .\qquad l=1,\cdots, s_1 \cr
&v_j^{(2)}=\nu_l .\qquad l=1,\cdots,(s-2 s_1) &\ecix \cr
}$$
with $\beta_l $ and $\nu_l $ both real.

Taking this in account, the equations \ecvi{a\hbox{-}b} result for the respective roots.
\eqna\ecx
$$\eqalignno{
&N_h \phi (2 \eta_k) -\sum_{i=1}^{s-2s_1}{\phi (2 \eta_k-2 \nu_i)} -
\sum_{i=1}^{s_1}{\phi ( \eta_k- \beta_i)}= 2 \pi \I_k^{(1)}
&\ecx a\cr
&N_h \phi ( \epsilon_k) -\sum_{i=1}^{s-2s_1}{\phi (\epsilon_k- \nu_i)} -
\sum_{i=1}^{s_1}{[\phi (2\epsilon_k- 2 \beta_i)+\phi ({2 \over 3}\epsilon_k- {2 \over 3} \beta_i)}]= 2 \pi \J_k^{(1)}
&\ecx b\cr
&N_p \phi (2 \nu_k) +\sum_{j=1}^{r-2r_1}{\phi (2\nu_k- 2\eta_j)} +\sum_{j=1}^{r_1}{\phi (\nu_k- \epsilon_j)}
-\sum_{i=1}^{s-2s_1}{\phi (\nu_k- \nu_i)}  -\cr
&\qquad \qquad \qquad\sum_{i=1}^{s_1}{[\phi (2\nu_k- 2 \beta_i)+  
\phi ({2 \over 3}\nu_k- {2 \over 3} \beta_i)}]= 2 \pi \I_k^{(2)}
&\ecx c\cr
&N_p \phi (\beta_k) +\sum_{j=1}^{r-2r_1}{\phi (\beta_k- \eta_j)} +
\sum_{j=1}^{r_1}{[\phi (2\beta_k- 2 \epsilon_j)+ 
\phi ({2 \over 3}\beta_k- {2 \over 3} \epsilon_j)}]- \cr
&\qquad \qquad \qquad\sum_{i=1}^{s-2 s_1}{[\phi (2\beta_k- 2 \nu_i)+  
\phi ({2 \over 3}\beta_k- {2 \over 3} \nu_i)}]- \cr
&\qquad \qquad \qquad\sum_{i=1}^{s_1}{[2 \phi (\beta_k-  \beta_i)+  
\phi ({1 \over 2}(\beta_k-  \beta_i))}]
= 2 \pi \J_k^{(2)}
&\ecx d\cr
}$$
where we have used the identities in appendix B.

In the thermodynamic limit the system is described in the language of  distribution functions of roots in both levels with particles and holes. In our case we define it as follows:
\eqnn\ecxi
$$\eqalignno{
&\rho_1=\lim_{N\rightarrow \infty}{{1 \over N(\eta_{j+1}-\eta_j)}}, \qquad
\sigma_1=\lim_{N\rightarrow \infty}{{1 \over N(\epsilon_{j+1}-\epsilon_j)}}, \cr
&\rho_2=\lim_{N\rightarrow \infty}{{1 \over N(\rho_{j+1}-\rho_j)}}, \qquad
\sigma_2=\lim_{N\rightarrow \infty}{{1 \over N(\beta_{j+1}-\beta_j)}}, &\ecxi\cr
}$$
and we call $[B_1]$, $[B_2]$, $[C_1]$ and $[C_2]$ the regions where  $\rho_1$ and $\rho_2$, $\sigma_1$ and $\sigma_2$  are defined respectively.

As usual\ref\rci{
C. N. Yang and C. P. Yang, {\it J. Math.Phys.} {\bf 54} (1969) 1115.  \hfill\break
V. E. Korepin, N. M. Bogoliubov and A. G. Izergin,{\it Quantum Inverse Scatering Method end Correlation Functions}, Cambridge: Cambridge University Press, (1993).}, we introduce the functions,
\eqna\ecxii
$$\eqalignno{
&Z_{\rho_1}(\lambda)={1 \over 2 \pi N}\biggl[  N_h \phi (2\lambda) -\sum_{i=1}^{s-2s_1}{\phi (2 \lambda - 2 \nu_i)} -
\sum_{i=1}^{s_1}{\phi ( \lambda- \beta_i)}\biggr]
&\ecxii a\cr
&Z_{\sigma_1}(\lambda)={1 \over 2 \pi N}\biggl[  N_h \phi ( \lambda) -\sum_{i=1}^{s-2s_1}{\phi (\lambda- \nu_i)} -
\sum_{i=1}^{s_1}{[\phi (2\lambda- 2 \beta_i)+\phi ({2 \over 3}\lambda- {2 \over 3} \beta_i)}]\biggr]
&\ecxii b\cr
&Z_{\rho_2}(\lambda)={1 \over 2 \pi N}\biggl[  N_p \phi (2 \lambda) +\sum_{j=1}^{r-2r_1}{\phi (2\lambda- 2\eta_j)} +\sum_{j=1}^{r_1}{\phi (\lambda- \epsilon_j)}
-\sum_{i=1}^{s-2s_1}{\phi (\lambda- \nu_i)}  -\cr
&\qquad \qquad \qquad\sum_{i=1}^{s_1}{[\phi (2\lambda- 2 \beta_i)+  
\phi ({2 \over 3}\lambda- {2 \over 3} \beta_i)}] \biggr]
&\ecxii c\cr
&Z_{\sigma_2}(\lambda)={1 \over 2 \pi N}\biggl[  N_p \phi (\lambda) +\sum_{j=1}^{r-2r_1}{\phi (\lambda- \eta_j)} +
\sum_{j=1}^{r_1}{[\phi (2\lambda- 2 \epsilon_j)+ 
\phi ({2 \over 3}\lambda- {2 \over 3} \epsilon_j)}]- \cr
&\qquad \qquad \qquad\sum_{i=1}^{s-2 s_1}{[\phi (2\lambda- 2 \nu_i)+  
\phi ({2 \over 3}\lambda- {2 \over 3} \nu_i)}]- 
\sum_{i=1}^{s_1}{[2 \phi (\lambda-  \beta_i)+  
\phi ({1 \over 2}(\lambda-  \beta_i))}]
\biggr] .\cr
&\qquad  &\ecxii d\cr
}$$

These functions are monotonyally increasing in their respective definition regions $[B_{1/2}]$ and $[C_{1/2}]$, and their values are integers when $\lambda$ takes the value of a root. Besides, there are other values of $\lambda$ where the $Z$ functions take a integer or a half-odd integer value, but they do not correspond to a root. We call each of these values a hole. The distribution functions are the derivatives of the $Z$ functions \rci ,
\eqn\ecxiii{
{d \over d\lambda}{Z_i (\lambda)}
\approx {N \over N_{h /  p} }\rho_i (\lambda) + {1 \over N_{h / p}} \sum_{i=1}^{}{\delta (\lambda-\theta_i)}
}
where $\theta_i$ are the position of the corresponding holes. The  thermodynamic limit is obtained  by doing,
\eqn\ecxiv{
\lim_{N\rightarrow \infty}{{1 \over N}\sum_{i}^{}{f(\lambda_i)}}\approx\int{d \lambda f(\lambda) \rho(\lambda)}.
}
Then, from \ecxii{a\hbox{-}d}, the corresponding distribution functions in this limit  are,
\eqna\ecxv
$$\eqalignno{
&{N \over N_h} \rho_1(\lambda)=
{1 \over 2\pi} \biggl[ 2 \phi' (2\lambda)-  
{2N \over N_h}\int_{[B_2]}{\phi'(2\lambda-2\mu) \rho_2(\mu)  d\mu} \cr
&\qquad\qquad\qquad -{N \over N_h}
\int_{[C_2]}{\phi'(\lambda-\mu) \sigma_2(\mu)  d\mu} \biggr]
&\ecxv a\cr
&{N \over N_h}\sigma_1(\lambda)=
{1 \over 2\pi} \biggl[ 2 \phi' (\lambda)- {N \over N_h}\int^{}_{[B_2]}{\phi'(\lambda-\mu) \rho_2(\mu)  d\mu} \cr
&\qquad\qquad\qquad
-{2N \over N_h}
\int_{[C_2]}{[\phi'(2(\lambda-\mu)) +{1 \over 3}\phi'({2 \over 3}(\lambda-\mu)] \sigma_2(\mu) \ d\mu}  \biggr]
&\ecxv b\cr
&{N \over N_p} \rho_2(\lambda)=
{1 \over 2\pi} \biggl[ 2 \phi' (2\lambda)+  
{2N \over N_p}\int_{[B_1]}{\phi'(2\lambda-2\mu) \rho_1(\mu)  d\mu} \cr
&\qquad\qquad\qquad +{N \over N_p}
\int_{[C_1]}{\phi'(\lambda-\mu) \sigma_1(\mu) d\mu} -
 {N \over N_p}\int_{[B_2]}{\phi'(\lambda-\mu)\rho_2(\mu) d\mu} \cr
&\qquad\qquad\qquad
-{2N \over N_p} 
\int_{[C_2]}{[\phi'(2(\lambda-\mu)) +{1 \over 3}\phi'({2 \over 3}(\lambda-\mu)] \sigma_2(\mu) \ d\mu}
 \biggr]
&\ecxv c\cr
&{N \over N_h}\sigma_2(\lambda)=
{1 \over 2\pi} \biggl[  \phi' (\lambda)+ {N \over N_p}\int^{}_{[B_1]}{\phi'(\lambda-\mu) \rho_1(\mu)  d\mu} \cr
&\qquad\qquad\qquad
+{2N \over N_p}
\int_{[C_1]}{[\phi'(2(\lambda-\mu)) +{1 \over 3}\phi'({2 \over 3}(\lambda-\mu)] \sigma_1(\mu) \ d\mu} \cr
&\qquad\qquad\qquad
+{2N \over N_p}
\int_{[B_2]}{[\phi'(2(\lambda-\mu)) +{1 \over 3}\phi'({2 \over 3}(\lambda-\mu)] \rho_2 (\mu) \ d\mu}  \cr
&\qquad\qquad\qquad
-{N \over N_p}
\int_{[C_1]}{[ 2 \phi'(\lambda-\mu) +{1 \over 2}\phi'({1 \over 2}(\lambda-\mu))]\sigma_2(\mu) \ d\mu} 
\biggr]
&\ecxv d\cr
}$$

Quantities corresponding to observables defined before,  can be found on this state in the thermodynamic limit in the same form, so 
the energy is given by
\eqn\ecxvi{
{E \over N}= J \biggl[ {-N_h+ 2 N_p \over N}+2 \int_{[B_1]}{\phi'(2\mu )\rho_1 (\mu) \ d\mu} +\int_{[C_1]}{\phi'(\mu )\sigma_1 (\mu) \ d\mu}\biggr].
}
The $O_1$ index (number of electrons) defined in \eci {a} is given by
\eqn\ecxvii{
n_e \equiv ={O_1 \over N}={r+N_p \over N}={N_p \over N}+
\int_{[B_1]}{\rho_1 (\mu) \ d\mu} +
\int_{[C_1]}{\sigma_1 (\mu) \ d\mu}.
}
The difference of magnetization $S^z$ between the $t\hbox{-}J$ sites minus the magnetization in the $J\hbox{-}t$ sites \eci {c,d},
\eqnn\ecxviii
$$\eqalignno{
s^z={S^z_{t-J} - S^z_{J-t} \over N}&={O_3-O_4 \over 2 N}= { 1\over 2N}(r+N_p-2 s) \cr
&={n_e \over 2}- \int_{[B_2]}{\rho_2 (\mu) \ d\mu}
-2 \int_{[C_2]}{\sigma_2 (\mu) \ d\mu}. &\ecxviii \cr
}$$
Using \ecxv{b,c} and \ecxvi, we find an important relation,
\eqn\ecxix{
{E \over N}=J \biggl[ {N_h- 2 N_p \over N} - 2\pi \bigl(\sigma_1(0)-\rho_2(0) \bigr)\biggr]
}

In our model, as we said before, we are going to suppose that the configuration of the ground state is given by two-string roots in both levels of the BAE, as it is suggested for the $t\hbox{-}J$ model in ref. \rvib . With this hypothesis, since $\rho_{1}=\rho_{2}=0$, the equations \ecxv{a\hbox{-}d} are reduced to two equations and are simplified considerably,
\eqna\ecxx
$$\eqalignno{
&{N \over N_h}\sigma_1(\lambda)=
{1 \over 2\pi} \biggl[ 2 \phi' (\lambda)-{2N \over N_h}
\int_{[C_2]}{[\phi'(2(\lambda-\mu)) +{1 \over 3}\phi'({2 \over 3}(\lambda-\mu)] \sigma_2(\mu) \ d\mu}  \biggr]
&\ecxx a\cr
&{N \over N_h}\sigma_2(\lambda)=
{1 \over 2\pi} \biggl[  \phi' (\lambda)+{2N \over N_p}
\int_{[C_1]}{[\phi'(2(\lambda-\mu)) +{1 \over 3}\phi'({2 \over 3}(\lambda-\mu)] \sigma_1(\mu) \ d\mu} \cr
&\qquad\qquad\qquad
-{N \over N_p}
\int_{[C_1]}{[ 2 \phi'(\lambda-\mu) +{1 \over 2}\phi'({1 \over 2}(\lambda-\mu))]\sigma_2(\mu) \ d\mu} 
\biggr]
&\ecxx b\cr
}$$
and the integration regions can be determined by $n_e$ and $s^z$.
\eqna\ecxxi
$$\eqalignno{
&n_e = {N_p \over N}+ 2
\int_{[C_1]}{\sigma_1 (\mu) \ d\mu}. &\ecxxi a\cr
&s^z={n_e \over 2}-2 \int_{[C_2]}{\sigma_2 (\mu) \ d\mu}. &\ecxxi b\cr
}$$
These equations can be solved numerically for any filling $n_e$  and magnetization
 $s^z$. 

If we take a system where the integration limits turn out to be,
$$
[C_1]= [C_2]=(-\infty, \infty),
$$
our equations can be solved analytically using the Fourier transform and some identities in Appendix B.

Let be
\eqn\ecxxii{
\sigma_j(\lambda)=
\int_{-\infty}^{\infty}{\hat{\sigma}_j(\alpha) e^{i \alpha\lambda} d\alpha}.
}
then, we obtain,
\eqna\ecxxiii
$$\eqalignno{
&\hat{\sigma}_1(\alpha)={ N_h \over N}{{e^{{-|\alpha| / 2} }  }\over{2 \cosh{{\alpha \over 2}}        }   } -
{ N_p \over N}{{e^{{-|\alpha| / 2} }  }\over{4 \cosh^2{{\alpha \over 2}}        }   }
&\ecxxiii a\cr
&\hat{\sigma}_2(\alpha)={ N_h \over N}{{e^{{-|\alpha| / 2} }  }\over{4 \cosh^2{{\alpha \over 2}}        }   } +
{ N_p \over N}{{e^{{|\alpha| / 2} }  }\over{8 \cosh^3{{\alpha \over 2}}        }   }
&\ecxxiii b\cr
}$$
These expressions can be used to determinate the main parameters of  our system
\eqna\eccxxiii
$$\eqalignno{
&{r \over N}=2 \int{\sigma_1 (\alpha) d\alpha}=2 \hat{\sigma}_1(0)={2 N_h-N_p \over 2 N} &\eccxxiii a\cr
&{s \over N}=2 \int{\sigma_2 (\alpha) d\alpha}=2 \hat{\sigma}_2(0)={2 N_h+N_p \over 4 N} &\eccxxiii b\cr
}$$
and using \ecxxi{a\hbox{-}b} we find,
\eqna\ecxxiv
$$\eqalignno{
&n_e= {2 N_h +N_p \over 2 N}, &\ecxxiv a\cr
&s^z=0,&\ecxxiv b\cr
}$$
We must note that for $N_p=0$, we have $n_e=1$ and  $ s^z=0$, this constitutes a antiferromagnetic state with half\hbox{-}filling.

From \ecxvi, the energy is given by
\eqn\ecxxv{
{E \over N}= J \bigl[ {N_h \over N} (1-2 \ln{2}) + { N_p\over N}({3 \pi \over 2}-3) \big]
}
and, for $N_p=0$ we obtain the result of the $t\hbox{-}J$ model.

As we know, it must verify
\eqn\ecxxvi{
{ N_p\over N} \leq n_e \leq      { N_h+2 N_p\over N}
}
then, from \ecxxiv{a}, this solution will only be true if,
\eqn\ecxxvii{
{ N_p\over N_h} \leq 2
}

\newsec{The excitation spectrum}
We start again from the BAE \ecrv{a,b}. 
 The more general string solutions will be of the form, 
\eqna\edvi
$$\eqalignno{
&v_{k,(m)}^{(1)} = v_{k,M}^{(1)} +i m, \qquad m=-M, \cdots,M &\edvi a\cr
&v_{k,(m)}^{(2)} = v_{k,M'}^{(2)} +i m, \qquad m=-M', \cdots,M' &\edvi b\cr
}$$
with $M$ and $M'$ integer or half-odd integer.

Following the same method as before, we multiply the equations of the same string, and we obtain that their center  $v_{k,M'}^{(i)}, i=1,2$, verifies the equations
\eqna\edvii
$$\eqalignno{
&2 N_h \arctan{{v_{k,M}^{(1)} \over M+ {1 \over 2} }}=
2 \pi I^{(1)}_{k,M}+\sum_{M''}^{}{\sum_{j=1}^{\nu_{M''}^{(2)}}{\Phi_{M,M''}
(v_{k,M}^{(1)}-v_{j,M''}^{(2)})} }
&\edvii a \cr
&-2 N_p \arctan{{v_{k,M}^{(2)} \over M+ {1 \over 2} }}=
-2 \pi I^{(2)}_{k,M}-\sum_{M'}^{}{\sum_{j=1}^{\nu_{M'}^{(2)}}{\Psi_{M,M'}
(v_{k,M}^{(2)}-v_{j,M'}^{(2)})} } \cr
&\qquad \qquad\qquad\qquad+\sum_{M''}^{}{\sum_{l=1}^{\nu_{M''}^{(1)}}{\Phi_{M,M''}
(v_{k,M}^{(2)}-v_{l,M''}^{(1)})} }
&\edvii b \cr
}$$
where $\nu_M^{(i)}$ is the number of strings in the $i$ level, and the $\Phi$ and $\Psi$ functions are defined in appendix B. A solution of \edvii{a,b}, is determinate by specifying  sets of  integers or half\hbox{-}odd integers $\{I^{(1)}_{k,M}\}$ and  the regions where the $\{C\}$  roots are distributed.

As we did before, we define the functions,
\eqna\edviii
$$\eqalignno{
&F_M^{(1)}(\lambda)={N_h \over \pi} \arctan{{\lambda \over M+ {1 \over 2} }}-
{1 \over 2\pi} \sum_{M''}^{}{\sum_{j=1}^{\nu_{M''}^{(2)}}{\Phi_{M,M''}
(\lambda-v_{j,M''}^{(2)})} } 
&\edviii a\cr
&F_M^{(2)}(\lambda)={N_p \over \pi} \arctan{{\lambda \over M+ {1 \over 2} }}-
{1 \over 2\pi} \sum_{M'}^{}{\sum_{j=1}^{\nu_{M'}^{(2)}}{\Psi_{M,M'}
(\lambda-v_{j,M'}^{(2)})} }\cr
&\qquad \qquad\qquad\qquad+{1 \over 2\pi} \sum_{M''}^{}{\sum_{j=1}^{\nu_{M''}^{(1)}}{\Phi_{M,M''}
(\lambda-v_{j,M''}^{(1)})} }, 
&\edviii b \cr
}$$
that are monotonic increasing and reach integer or half\hbox{-}odd integers values when $\lambda$  takes the value of a root.

By counting the number of  roots, and calling $H_M^{(i)}$ the number or holes in the sea of $M$\hbox{-}strings at level $i$, we have,
\eqn\edviv{
2 I^{(i)}_{max,M} +1 = \nu_M^{(i)}+ H_M^{(i)}
}
and supposing that the centers of the strings are distributed along the real numbers,
\eqn\edvv{
2 I^{(i)}_{max,M} +1 = F_M^{(i)}(\infty)
}
and then, using \edviii{a,b}, \edviv\  and \edvv, we obtain,
\eqna\edvvi
$$\eqalignno{
&\nu_M^{(1)}+H_M^{(1)}=N_h -2 \sum_{M''\geq 0}^{}{K(M,M'') \nu_{M''}^{(2)}}
&\edvvi a\cr
&\nu_M^{(2)}+H_M^{(2)}=N_p -2 \sum_{M'\geq 0}^{}{J(M,M') \nu_{M'}^{(2)}}+
2 \sum_{M''\geq 0}^{}{K(M,M'') \nu_{M''}^{(1)}}
&\edvvi b\cr
}$$
where
\eqn\edvvii
{
J(M_1,M_2)=\cases{
2M_1 +{1 \over 2} & if $ M_1=M_2$ , \cr
2 {\rm min} (M_1, M_2) +1 & if $ M_1\neq M_2$ , \cr}
}
and
\eqn\edvviii
{
K(M_1,M_2)=\cases{
M_2 +{1 \over 2} & if $ M_2+{1 \over 2} \leq M_1$ , \cr
 M_1 +{1 \over 2} & if $ M_2+{1 \over 2}>M_1$ , \cr}
}

Besides, the number of roots, obviously, must verify ,
\eqna\edvix
$$\eqalignno{
&r=\sum_{M\geq 0}^{}{(2 M+1) \nu_M^{(1)}} &\edvix a\cr
&s=\sum_{M\geq 0}^{}{(2 M+1) \nu_M^{(2)}} &\edvix a\cr
}$$
and now, we can apply \edvvi{a, b} and \edvix{a,b} to the ground state, according to which type of string we considered that forms it.

If we suppose, as before, that the ground state is formed with two sea of two-strings, we have,
\eqna\edvx
$$\eqalignno{
&H^{(1)}_{{1 \over 2}}=N_h-\nu^{(1)}_{{1 \over 2}}-\nu^{(2)}_{0}-2\sum_{M''\geq {1 \over 2}}^{}{\nu^{(2)}_{M''}} &\edvx a \cr
&H^{(2)}_{{1 \over 2}}=N_p- 2 \nu^{(2)}_{0}  -4\sum_{M'\geq {1 \over 2}}^{}{\nu^{(2)}_{M'}}   -\nu^{(1)}_{0}-2\sum_{M''\geq {1 \over 2}}{\nu^{(1)}_{M''}} &\edvx b \cr
}$$
and then, with this results, we obtain from \eccxxiii{a,b},
\eqn\edvxi{
H^{(1)}_{{1 \over 2}}=H^{(2)}_{{1 \over 2}}=0,
}
that is to say, we have two two-string seas without holes.

Under these hypothesis we can analyze  different types of excitations by introducing holes in the ground state and keeping constant some observables. The one that we are going to consider is maintaining constant the electron number and the magnetization
\eqn\edvxii{
n_e=\hbox{\rm const.}, \qquad s_z=\hbox{\rm const. },
}
then, by imposing \edvix{a,b} and \edvx{a,b}, we obtain that the state must be characterized having two real roots and a hole in level $(1)$
\eqn\edvxiii{
\nu^{(1)}_{0}=2, \qquad  H^{(1)}_{{1 \over 2}}=1,
}
and besides,
\eqn\edvxiv{
\matrix{
\nu^{(2)}_{0}=0, \quad & \nu^{(1)}_{{1 \over 2}}={2N_h-N_p \over 4} -1 ,\quad& \nu^{(1)}_{M\geq{ 1\over 2}} =0,\cr
& & \cr
H^{(2)}_{{1 \over 2}}=0 , \quad& \nu^{(2)}_{{1 \over 2}}={2N_h+N_p \over 8}, \quad & \nu^{(2)}_{M\geq{ 1\over 2}} =0,\cr
}.
}
These conditions correspond to one the two states, the first one is obtain from the ground state by changing two consecutive sites with spin up to a new state with the two sites in spin down, the second one is a state with one of the two sides without electrons and the other with a pair.

Under our hypothesis about the ground state, we can calculate the contribution of  every hole and real root in both levels, to the energy of a excited state compared with the energy of the ground state.
Using \ecxiii ,
the equations \ecxx{a,b} 
must change to,
\eqna\edvxv
$$\eqalignno{
&{N \over N_h}\sigma_1(\lambda)+{1 \over N_h} \sum_{h_1}^{}{\delta(\lambda-\theta_{h_1}})=
{1 \over 2\pi} \biggl[  \phi' (\lambda)-
{1 \over N_h} \sum_{r_2}^{}{\phi'(\lambda-\theta_{r_2}}) \cr
& \qquad \qquad\qquad
-{2N \over N_h}\int_{[C_2]}{[\phi'(2(\lambda-\mu)) +{1 \over 3}\phi'({2 \over 3}(\lambda-\mu)] \sigma_2(\mu) \ d\mu}  \biggr]
&\edvxv a\cr
&{N \over N_p}\sigma_2(\lambda)
+{1 \over N_p} \sum_{h_2}^{}{\delta(\lambda-\theta_{h_2}})=
{1 \over 2\pi} \biggl[  \phi' (\lambda)+
{1 \over N_p} \sum_{r_1}^{}{\phi'(\lambda-\theta_{r_1}})
\cr
& \qquad \qquad
 -{2 \over N_p} \sum_{r_2}^{}{[\phi'(2(\lambda-\theta_{r_2})) +{1 \over 3}\phi'({2 \over 3}(\lambda-\theta_{r_2})] }
\cr
&\qquad\qquad
+{2N \over N_p}\int_{[C_1]}{[\phi'(2(\lambda-\mu)) +{1 \over 3}\phi'({2 \over 3}(\lambda-\mu)] \sigma_1(\mu) \ d\mu} \cr
&\qquad\qquad
-{N \over N_p}
\int_{[C_1]}{[ 2 \phi'(\lambda-\mu) +{1 \over 2}\phi'({1 \over 2}(\lambda-\mu))]\sigma_2(\mu) \ d\mu} 
\biggr]
&\edvxv b\cr
}$$
where $\theta_{h_i}$ parametrizes the holes at level ${i}$ and  $\theta_{r_i}$ the real roots.

The system can be solved as before, by using the Fourier transform, and we obtain that the distribution functions can be written as,
\eqna\edvxvi
$$\eqalignno{
&\hat{\sigma}_1(\alpha)=\hat{\sigma}_{1 (0)}(\alpha)+\hat{\sigma}_{1 (n)}(\alpha)
&\edvxvi a\cr
&\hat{\sigma}_2(\alpha)=\hat{\sigma}_{2 (0)}(\alpha)+\hat{\sigma}_{2 (n)}(\alpha)
&\edvxvi b\cr
}$$
with 
\eqna\edvxvii
$$\eqalignno{
&\hat{\sigma}_{1 (0)}(\alpha)=
{ N_h \over N}{{e^{{-|\alpha| } }  }\over{1+ e^{-|\alpha |}        }   } -
{ N_p \over N}{{e^{{-3 |\alpha| /2} }  }\over{(1+ e^{-|\alpha |} )^2       }   },
&\edvxvii a \cr
&\hat{\sigma}_{2 (0)}(\alpha)=
{ N_h \over N}{{e^{{-3|\alpha| /2} }  }\over{(1+ e^{-|\alpha |})^2        }   } +
{ N_p \over N}{{e^{{- |\alpha| } }  }\over{(1+ e^{-|\alpha |} )^3       }   },
&\edvxvii b \cr
&\hat{\sigma}_{1 (n)}(\alpha)={ 1 \over N} \bigg[
\sum_{h_2}^{}{  { e^{-|\alpha| /2 } \over (1+ e^{-|\alpha |} )^2 }  e^{-i\alpha \theta_{h_2}}  }
-\sum_{h_1}^{}{  { 1 \over (1+ e^{-|\alpha |} ) }  e^{-i\alpha \theta_{h_1}}  } \cr
&\qquad\qquad\qquad
-\sum_{r_1}^{}{  { e^{-3|\alpha| /2 } \over (1+ e^{-|\alpha |} )^2 }  e^{-i\alpha \theta_{r_1}}  }      \bigg],
&\edvxvii c \cr
&\hat{\sigma}_{2 (n)}(\alpha)={ 1 \over N} \bigg[
\sum_{h_1}^{}{  { e^{-|\alpha| /2 } \over (1+ e^{-|\alpha |} )^2 }  e^{-i\alpha \theta_{h_1}}  }
-\sum_{h_2}^{}{  { 1 \over (1+ e^{-|\alpha |} )^3 }  e^{-i\alpha \theta_{h_2}}  } \cr
&\qquad\qquad\qquad
+\sum_{r_1}^{}{  { e^{-|\alpha|  } \over (1+ e^{-|\alpha |} )^3 }  e^{-i\alpha \theta_{r_1}}  } +
 \sum_{r_2}^{}{  { e^{-|\alpha|/2  } \over (1+ e^{-|\alpha |} )}  e^{-i\alpha \theta_{r_2}}  }    \bigg].
&\edvxvii d \cr
}$$

The contribution to the energy per site is,
\eqn\edvxviii{
\Delta e\equiv {\Delta E \over N}= \int{\phi'(\mu)\sigma_{1 (n)} (\mu) d\mu +{1 \over N}\sum_{r_1}^{}{{1 \over (\theta^2_{r_1}+{1 \over 4})}}}.
}
From this expression, we observe that the holes in both levels and only the real roots of the first level give contribution to the energy. A straighforward calculation show the following rules:

 i.- every hole in the first level gives a contribution  $\Delta e=-\varepsilon_1(\theta_{h_1})$

 ii.- every hole in the second level gives a contribution  $\Delta e=\varepsilon_2(\theta_{h_2})$

 iii.- every real root in the first level gives a contribution  
$$\Delta e={1 \over N}\sech (\pi \theta_{r_1})+\varepsilon_2 (\theta_{r_1})$$

iv.- every real root in the second level gives a contribution   $\Delta e =0$,

\noindent being
\eqn\edvxix{
\varepsilon_n (v)={1 \over n N}\int_{0}^{\infty}{ \cos{\alpha v}\over e^{{ \alpha\over 2}} \cosh^n{{\alpha \over 2}}} d\alpha.
}

We can apply these rules to the first example, where we have held constant the number of electrons and magnetization respect to the ground state and it is characterized by two real roots and a hole.  We can suppose that the roots and the hole are parametrized by the same value $\theta$, it is said, the state is coming by changing a two-string of the ground state by two real roots, which leave a hole in the two-string sea. Doing the integrals in $\Delta e$, we obtain for the state,
\eqnn\edvxvii
$$\eqalignno{
&\Delta e={1 \over N} \bigg[
{2 \over \cosh{(\pi \theta)}} +2 \Re[\Psi({1 \over 2}+i \theta)]- 2 \Re[\Psi({1 \over 4}+i {\theta \over 2})] \cr
&\qquad\qquad\qquad\qquad
- 4\theta \big(\Im[\Psi({1 \over 2}+i \theta)]-  \Im[\Psi({1 \over 4}+i {\theta \over 2})]\big) \cr
&\qquad\qquad\qquad\qquad
-2 \Re[ \Psi(i{\theta \over 2})+2 \Re[\Psi(i \theta)] -2 ( 1+2\ln{2)}\bigg]
&\edvxvii\cr
}$$ 
where $\Psi(x)$ is the derivative of the logarithm of the Euler gamma function.

Taking an alternating chain $N_h=N_p=N/2$,  we can represent the energy given by \edvxvii. The results are shown in fig.1. As we can see, $\Delta e$ is null out of an interval around the origin. Then a conclusion is that there is not energy gap between the excited and the ground state.


{ \tenbf  Acknowledgements}

We are grateful to J. I. Abad for the careful reading of the manuscript.
This work was partially supported by the Direcci\'{o}n General de Ense\~{n}anza Superior, Grant No PB97-0738.

\appendix {A} {}

Using the relation of commutation \ebv c\ we have

\eqnn\eapi
$$\eqalignno{
&C_{a_{1}}(\lambda_{1}) C_{a_{2}}(\lambda_{2}) \ldots 
	 C_{a_{r}}(\lambda_{r}) =C_{b_{2}}(\lambda_{2})
	 C_{b_{3}}(\lambda_{3}) \ldots 
	  \qquad \qquad \cr
& \qquad \qquad \qquad \ldots C_{b_{r}}(\lambda_{r})
	  C_{b_{1}}(\lambda_{1})
	 G(\lambda_{1},\vec{\lambda})^{b_{1},\ldots,b_{r}}_{a_{1}, 
	 \ldots,a_{r}} \, , 
	&\eapi\cr
}$$
with
\eqn\eapii{
	G(u,\vec{\lambda})^{b_{1},\ldots,b_{r}}_{a_{1}, 
	 \ldots,a_{r}} = r(u-\lambda_{1})^{i_{1},a_{1}}_{b_{1},a}
	 r(u-\lambda_{2})^{i_{2},a_{2}}_{b_{2},i_{1}} \ldots
	 r(u-\lambda_{r})^{a,a_{r}}_{b_{r},i_{r-1}}\,.
}
Taking
\eqn\eapiii{
	M^{(j)}(\lambda_{j}) = G(\lambda_{j},\vec{\lambda}) \cdot
	G(\lambda_{j-1},\vec{\lambda}) \cdot \ldots \cdot
	G(\lambda_{1},\vec{\lambda}) \,,
}
we have the relation
\eqnn\eapiv
$$\eqalignno{
&C_{a_{1}}(\lambda_{1}) C_{a_{2}}(\lambda_{2}) \ldots 
	 C_{a_{r}}(\lambda_{r}) = C_{b_{k}}(\lambda_{k}) \ldots
	 C_{b_{r}}(\lambda_{r}) C_{b_{1}}(\lambda_{1}) \ldots 
	  \qquad \qquad \cr
& \qquad \qquad \qquad \ldots C_{b_{k-1}}(\lambda_{k-1}) 
	 	M^{(k-1)}(\lambda_{k-1})^{b_{1},\ldots,b_{r}}_{a_{1}, 
	 \ldots,a_{r}} \, . 
	&\eapiv\cr
}$$

\appendix{B} {}

In this appendix, we are going to give brief remarks about the function,
\eqn\aci{
\phi (x)\equiv 2 \arctan{(x)} =- i\ln{{1+i x \over 1-i x}},
}
that takes values in the interval $-\pi$ to $+\pi$ when $x $
goes from $-\infty$ to $\infty$,

The function has  the following properties, that can be proof by a straighforward calculation:
\eqn\acii{
\phi (x+i) + \phi (x-i) = \pi + \phi ({x \over 2})
}

\eqn\aciii{
\phi (x+2 i) + \phi (x-2 i) = \phi ({x \over 3}) - \phi (x) 
}

\eqn\aciv{
\phi (x+{i \over 2}) + \phi (x- {i \over 2}) = \phi ({2x \over 3})+ \phi (2x) 
}

The Fourier transform that we have used 
\eqn\aciv{
\rho(\lambda)={1 \over 2 \pi}\int_{-\infty}^{\infty}{\hat{\rho}(\alpha) e^ {i \alpha\lambda} d\alpha}, \quad\
\hat{\rho}(\alpha)=\int_{-\infty}^{\infty}{\rho(\lambda) e^ {-i \alpha\lambda} d\lambda},
}
The derivative of function $\phi$ is
\eqn\acv{
{d{\phi(\lambda)} \over dx}= {2 \over 1+\lambda^2}
}
and then, the Fourier transform,
\eqn\acvi{
\int_{-\infty}^{\infty} {\phi'(\lambda) e^ {-i \alpha\lambda} d\lambda} = e^{-|\alpha|}
}

It is convenient to define the functions
\eqn\acvii{
\Psi_{M_1,M_2}(x)=2 \sum_{n=|M_2 -M_1|}^{M_2 +M_1}{\big[\arctan{{x \over n}}+
\arctan{{x \over n+1}}\big]}
}
and
\eqn\acviii{
\Phi_{M_1,M_2}(x)=2 \sum_{n= -M_1}^{M_1}{\arctan{{x+i n \over M_2+{1 \over 2}}}}
}
 \vfill
\eject

\listrefs
\bigskip
\centerline{\epsfxsize=8cm  \epsfbox{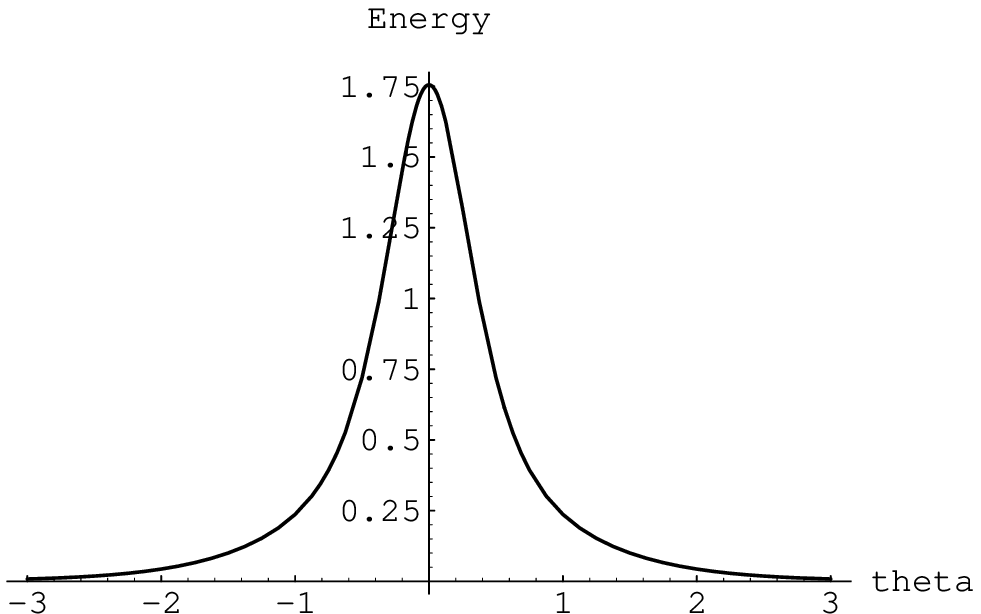}}
\centerline{Fig. 1}
\centerline { Energy of the Excited State}
\bigskip

\vfill
\eject

\end{document}